\documentclass[10pt,conference]{IEEEtran}
\IEEEoverridecommandlockouts
\usepackage{amsmath,amssymb,amsfonts}
\usepackage{algorithmic}
\usepackage{graphicx}
\usepackage{textcomp}
\usepackage{xcolor}
\usepackage{balance}
\usepackage{subcaption}
\usepackage{multirow}
\usepackage{booktabs}
\usepackage[nofillcomment,vlined,ruled,linesnumbered]{algorithm2e}
\def\BibTeX{{\rm B\kern-.05em{\sc i\kern-.025em b}\kern-.08em
    T\kern-.1667em\lower.7ex\hbox{E}\kern-.125emX}}

\newcommand{\name}{COSMos\xspace}
\newcommand{\cosmosfl}{COSMosFL\xspace}

\begin{document}

\title{\cosmosfl: Ensemble of Small Language Models for Fault Localisation\thanks{This work has been supported by the National Research Foundation of Korea (NRF) funded by the Korean government MSIT (RS-2023-00208998), and the Engineering Research Center Program funded by the Korean Government MSIT (RS-2021-NR060080).}}

\author{\IEEEauthorblockN{Hyunjoon Cho}
\IEEEauthorblockA{\textit{School of Computing} \\
\textit{KAIST}\\
Daejeon, Republic of Korea \\
hyunjoon.cho@kaist.ac.kr}
\and
\IEEEauthorblockN{Sungmin Kang}
\IEEEauthorblockA{\textit{School of Computing} \\
\textit{KAIST}\\
Daejeon, Republic of Korea \\
sungmin.kang@kaist.ac.kr}
\and
\IEEEauthorblockN{Gabin An}
\IEEEauthorblockA{\textit{School of Computing} \\
\textit{KAIST}\\
Daejeon, Republic of Korea \\
gabin.an@kaist.ac.kr}
\and
\IEEEauthorblockN{Shin Yoo}
\IEEEauthorblockA{\textit{School of Computing} \\
\textit{KAIST}\\
Daejeon, Republic of Korea \\
shin.yoo@kaist.ac.kr}
% \and
% \IEEEauthorblockN{Anonymous}
% \IEEEauthorblockA{\textit{ABC} \\
% \textit{ACME}\\
% City, Country \\
% email address or ORCID}
% \and
% \IEEEauthorblockN{Anonymous}
% \IEEEauthorblockA{\textit{ABC} \\
% \textit{ACME}\\
% City, Country \\
% email address or ORCID}
% \and
% \IEEEauthorblockN{Anonymous}
% \IEEEauthorblockA{\textit{ABC} \\
% \textit{ACME}\\
% City, Country \\
% email address or ORCID}
}

\maketitle

\begin{abstract}
LLMs are rapidly being adopted to build powerful tools and agents for software engineering, but most of them rely heavily on extremely large closed-source models. This, in turn, can hinder wider adoption due to security issues as well as financial cost and environmental impact. Recently, a number of open source Small Language Models (SLMs) are being released and gaining traction. While SLMs are smaller, more energy-efficient, and therefore easier to locally deploy, they tend to show worse performance when compared to larger closed LLMs. We present \name, a task-level LLM ensemble technique that uses voting mechanism, to provide a broader range of choice between SLMs and LLMs. We instantiate \name with an LLM-based Fault Localisation technique, AutoFL, and report the cost-benefit trade-off between LLM accuracy and various costs such as energy consumption, inference time, and the number of tokens used. An empirical evaluation using Defects4J shows that \name can build effective ensembles that can achieve Pareto-optimality in terms of FL accuracy and inference cost, when compared to individual models.
\end{abstract}


%!TEX root=../paper.tex

\section{Introduction}
\label{sec:introduction}

Large Language Models (LLMs) are rapidly being adopted by software engineers to automate various tasks across the software development lifecycle~\cite{Fan2023yu}. While LLMs are essentially autocompletion engines trained on a vast amount of data~\cite{Vaswani2017aa}, they have exhibited many useful emergent behaviour, including their capability to perform in-context learning. This has led to many advanced prompting/inference techniques, such as Chain-of-Thought~\cite{Wei2024aa}, self-consistency~\cite{wang2022self}, and ReAct~\cite{Yao2022qf}. Increasingly, these techniques are being used to build LLM-based \emph{agents}~\cite{Feldt2023ax,Bouzenia2024aa,Yoon2024aa}.

%In the meantime, most existing works are coupled with proprietary models, especially the GPT. This hinders the wider application of emerging techniques, depending on the environments.
One challenge in broader adoption of these techniques is the dependence on commercial and closed LLMs. In addition to the financial cost of using those models, organisations may not want to reveal their source code as part of any prompt that are fed to external LLMs, for security reasons. Finally, while growing LLM sizes have so far been accompanied with improving performance, there are concerns about the environmental impact that these large models have~\cite{Strubell2019aa,Rillig2023aa}. 

% These dependencies often restrict their applicability across diverse environments, hindering the practical integration and generalization of LLM-based techniques.

%Small language models (SLMs) are on a rise in terms of capability, thus as an alternative for the above problem. Yet individual capabilities are unmatched to that of GPT. Thus, we tried to tackle this problem by leveraging the power of multiple SLMs.
Recently, Small Language Models (SLMs) with open source licenses have started gaining traction due to their improving capabilities and ability to be hosted and served more locally, offering an alternative to the larger, closed models and their limitations~\cite{dubeyLlama3Herd2024,teamGemma2Improving2024,huiQwen25CoderTechnicalReport2024}. However, individual SLMs typically fall short of the comprehensive performance delivered by larger language models like GPT-4~\cite{achiam2023gpt}, creating a noticeable gap in their stand-alone effectiveness~\cite{Kang2024aa}. Consequently, a potential user of LLM-based software engineering technique is presented with two options: high performance with high cost, or low performance with low cost. It would be ideal to have a broader range of choices in the cost-benefit trade-off.

We propose \name (\textbf{CO}llection of \textbf{S}mall Language \textbf{Mo}del\textbf{s}), an ensemble of SLMs: it builds upon self-consistency, a prompting technique that states, for logical tasks, it is better to take multiple samples of LLM answers and marginalise them~\cite{wang2022self}. While self-consistency has been widely accepted as a simple yet effective validation technique that can improve the correctness of LLM-generated solutions~\cite{Kang2024aa,Ahmed2023aa,kangQuantitativeQualitativeEvaluation2024a}, it exacerbates the issue of cost due to its need to sample multiple LLM-generated solutions. \name aims to both exploit and alleviate challenges that self-consistency introduces. Instead of taking multiple samples from a closed LLM, \name forms an ensemble of SLMs and aggregate their answers to form the final solution. In turn, we hope to lower the cost of the self-consistency based inferences by using SLMs, while maintain high FL accuracy. 

In this paper, we concretely evaluate \name by instantiating it with ensembles of LLM-based Fault Localisation (FL) technique, AutoFL~\cite{kangQuantitativeQualitativeEvaluation2024a}, to obtain \cosmosfl. We first choose the membership of the ensemble based on the FL performance of individual SLMs. Subsequently, the ensemble of SLMs provide the multiple inference samples that result in the final ranking of the likely faulty methods via voting. We evaluate two ensemble schemes: one where each member SLM has the equal voting power, and another where we optimise the relative voting weights with the aim of improving the resulting FL accuracy. We report not only the FL accuracy of the ensemble, but also various cost measures including number of tokens, inference time, and the overall energy consumption. Our evaluation of \cosmosfl using Defects4J~\cite{just2014defects4j} shows that ensembles can indeed outperform individual models when the FL task is constrained by energy consumption or token count. To facilitate further research and reproducibility, we publicly release our implementation at GitHub\footnote{https://github.com/coinse/cosmosfl}. Detailed technical contributions are as follows:

\begin{itemize}
\item We introduce \name, an ensemble of multiple open source SLMs. While ensembles of LLMs have been proposed for token level decoding, \name is the first to introduce task-level voting based ensembles.

\item We compare two ensemble methods: a vanilla voting-based ensemble, and a weight-optimised ensemble. The weight optimisation uses Differential Evolution to optimise the weights that are applied to votes cast by membership SLMs.

\item We instantiate \name with an LLM-based FL technique, AutoFL, to obtain \cosmosfl. We evaluate its performance using the widely studied Defects4J benchmark. We report the trade-off between FL accuracy and various costs, such as power consumption, token size, and inference time.
\end{itemize}

The remainder of the paper is organised as follows. Section~\ref{sec:approach} presents AutoFL and the design of \name. Section~\ref{sec:experimentalsetup} presents the settings of the empirical evaluation, the results of which are shown in Section~\ref{sec:results}. Section~\ref{sec:discussion} discusses details of our findings. Section~\ref{sec:relatedwork} presents related work that are relevant to ours, and Section~\ref{sec:conclusion} finally concludes.

%!TEX root=../paper.tex

\section{Approach}
\label{sec:approach}

\subsection{Preliminaries}

\subsubsection{Self-Consistency} Self-consistency is a property of LLM-based reasoning that, for a complex reasoning task, taking multiple samples of LLM inferences and marginalising over them tend to yield more accurate answers when compared to greedy decoding~\cite{wang2022self}. This has been widely adopted by applications of LLMs for software engineering tasks~\cite{Ahmed2023aa,kangQuantitativeQualitativeEvaluation2024a}.

\subsubsection{AutoFL} AutoFL~\cite{kangQuantitativeQualitativeEvaluation2024a} is a LLM agent for repository-level FL, incorporating four tools to gather project-related information: 1) class-level coverage of the failing test, 2) method-level coverage within a covered class, 3) the code snippet of the given method, and 4) comments associated with the method. The original AutoFL utilised GPT-3.5 and GPT-4.

To improve accuracy and reliability, AutoFL leverages self-consistency~\cite{wang2022self} by performing FL $R$ times independently for a single bug. In each inference run, it predicts a set of likely buggy methods, and the results are aggregated using a voting-based mechanism. Specifically, each predicted method in a run is assigned a score equal to 1 divided by the total number of predicted methods in that run. These scores are averaged across all runs, ensuring that the sum of the scores for all methods equals 1, and are then used to derive the final ranking.
For example, if the predicted methods from five runs ($R=5$) are \(\{m_1, m_2\}\), \(\{m_2\}\), \(\{m_2\}\), \(\{m_2\}\), and \(\{m_3\}\), the score of \(m_2\) is \(\left(\frac{1}{2} + 1 + 1 + 1 + 0\right)/5 = 0.7\), while \(m_1\) and \(m_3\) get \(0.1\) and \(0.2\), respectively. In AutoFL, the confidence in its result is defined by the maximum score of the methods, e.g., \(0.7\) in the previous example: a higher maximum score indicates better alignment of results across multiple runs.

AutoFL demonstrates that LLMs can be effectively applied to FL when augmented with specialised tools to extract contextual information. However, its original evaluation is limited to GPT models, which are proprietary and commercially available LLMs. While GPT models generally exhibit strong performance, their use may not always be feasible due to concerns such as code security and monetary costs.

\subsection{Ensemble of Small Language Models}

\begin{figure*}[t]
\centerline{\includegraphics[width=0.9\textwidth]{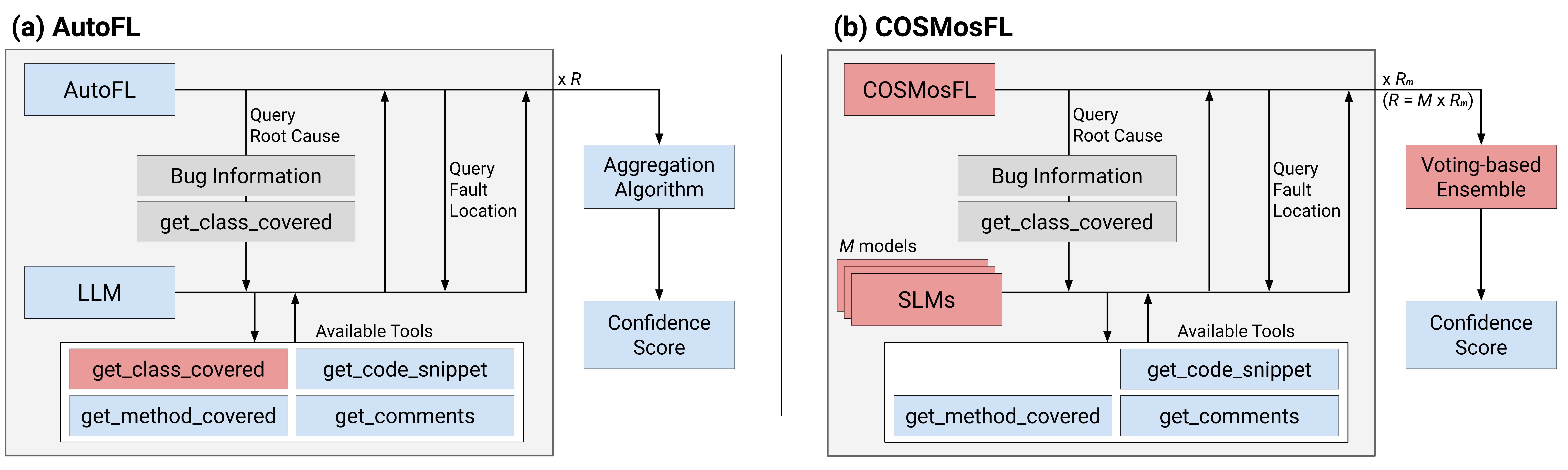}}
\caption{Overview of our approach against AutoFL~\cite{kangQuantitativeQualitativeEvaluation2024a} with differences colored in red.}
\label{fig:overview}
\end{figure*}

Intuitively, \name is a voting-based ensemble of SLMs that makes use of the multiplicity of reasoning samples required by self-consistency. Ensembles of SLMs can be constructed by collecting multiple reasoning samples from participating SLMs, instead of repeated sampling of a single LLM. Subsequently, \name marginalises over the samples using the voting-based ensemble mechanism. We posit that \name can be applied to any task for which self-consistency is shown to improve the performance of LLMs.

We instantiate \name with Fault Localisation (FL) problem and propose COSMosFL (\textbf{CO}llection of \textbf{S}mall Language \textbf{Mo}del\textbf{s} for FL), a novel approach that enables the application of diverse SLMs for FL (the overview of \cosmosfl is shown in Figure~\ref{fig:overview}, along with AutoFL). We evaluate whether an ensemble of heterogeneous open-source SLMs, capable of running on local machines without network access (ensuring high security), can also effectively address the FL problem. Further, \cosmosfl will allow us to investigate the cost-benefit trade-offs of our ensemble approach, \name, with respect to a concrete task.

% We propose COSMosFL(\textbf{CO}llection of \textbf{S}mall Lanague \textbf{Mo}del\textbf{s} for FL) to enable application of diverse SLMs for the FL problem. Figure \ref{fig:overview} illustrates the overview of our technique, highlighting the difference between the original work and ours with red boxes.

To explore the effectiveness of SLMs in performing the FL task within the AutoFL framework, we first replaced GPT with open-source models—Llama3 8B~\cite{dubeyLlama3Herd2024} and Gemma2 9B~\cite{teamGemma2Improving2024}. Our initial experiments found that smaller models frequently called the first tool (class-level coverage) redundantly, diminishing performance. Based on this observation, we remove the class-level coverage tool for \cosmosfl. 
%Table \ref{tab:performance_comparison} demonstrates an improvement in FL performance from the template modification.

% \begin{table}[t]
% \caption{Performance Comparison on Defects4J~\cite{just2014defects4j} (353 Bugs)}
% \begin{center}
% \resizebox{\columnwidth}{!}{%
% \begin{tabular}{llrrrrr}
% \toprule
% \multirow{2}{*}{\textbf{Model}} & \multirow{2}{*}{\textbf{Template}} & \multicolumn{5}{c}{\textbf{Accuracy (acc@k)}} \\
% \cmidrule{3-7}
% & & \textbf{\textit{acc@1}} & \textbf{\textit{acc@2}} & \textbf{\textit{acc@3}} & \textbf{\textit{acc@4}} & \textbf{\textit{acc@5}} \\ \midrule
% Llama3 8B & original &  80 & 115 & 131 & 140 & 148 \\
% Llama3 8B & modified & \textbf{108} & \textbf{147} & \textbf{168} & \textbf{180} & \textbf{190} \\ \midrule
% Gemma2 9B & original & 105 & 134 & 150 & 160 & 166 \\
% Gemma2 9B & modified & \textbf{112} & \textbf{145} & \textbf{159} & \textbf{169} & \textbf{182} \\
% \bottomrule
% \end{tabular}}
% \label{tab:performance_comparison}
% \end{center}
% \end{table}

% \begin{figure}[t]
% \centerline{\includegraphics[width=0.65\columnwidth]{EOL/figs/approach_overlap.pdf}}
% \caption{Overlap of top-ranked bugs identified by Llama3 8B (left) and Gemma2 9B (right) over five individual repetitions.}
% \label{fig:appraoch_overlap}
% \end{figure}

The initial evaluation of two models also show that they rank distinct sets of bugs at the top, a characteristic we refer to as orthogonality. Comparing bugs ranked at the top by Llama3 8B and Gemma2 9B over five repetitions, we see that 71 bugs are ranked at the top by both models, while 37 exclusively by Llama3 8B and 41 exclusively by Gemma2 9B: approximately 35\% of correctly localised bugs are uniquely identified by each model. This orthogonality suggests that individual models may excel at localizing certain types of faults missed by others, providing a foundation for an ensemble of heterogeneous SLMs to enhance FL performance by leveraging their complementary behaviour. 
% This motivates us to explore ensembles with a wider range of SLMs.

Building on the confidence scores generated by AutoFL, we develop a voting-based ensemble technique. Since the original AutoFL already uses voting as the aggregation mechanism, it is straightforward to implement a voting-based ensemble: instead of $R$ inference runs from a single LLM, \cosmosfl takes $R_M$ inference runs per each of the $M$ member models of the ensemble, such that $R_M \times M = R$. Subsequently, the same voting-based aggregation takes place, producing the final confidence score based on the ensemble of SLMs.

% This approach aggregates rankings by calculating a weighted sum of the confidence scores from each model, producing a combined ranking that leverages the strengths of individual models. Unlike the original use of confidence scores in AutoFL, which focuses only on the maximum score, we extend the concept to encompass all methods associated with a bug so that the information collected during individual executions is fully utilized.

\SetKwComment{Comment}{/* }{ */}
\begin{algorithm}
\caption{Differential Evolution\label{alg:de}}
\KwIn{Problem Dimension $n$, Fitness Evaluator \textbf{fitness}}
\KwOut{Best Agent $best$}

$n_{pop} \gets$ Population Size\;
$n_{gen} \gets$ Number of Generations\;
$p_{cx} \gets$ Crossover Probability\;
$w_{d} \gets$ Differential Weight\;
$pop \gets$ randomly generate $n_{pop}$ agents\;

$gen \gets 0$\;
\While{$gen < n_{gen}$} {
    \For{$agent$ in $pop$}{
        $agent_{ref} \gets$ \textbf{clone}($agent$)\;
        $a, b, c \gets$ \textbf{sample\_three\_from}($pop$)\;
        $R \gets$ \textbf{random}(\textit{[1, ..., $n$]})\;
        \For{$i$ in [1, ..., $n$]}{
            \If{$i = R$ or \textbf{uniform}(0, 1) $<$ $p_{cx}$} {
                $agent_{ref}[i] \gets a[i] + w_{d} * (b[i] - c[i])$\;
            }
        }
        \If{\textbf{fitness}($agent$) $<$ \textbf{fitness}($agent_{ref}$)} {
            $agent \gets agent_{ref}$\;
        }
    }
    $best_{gen} \gets$ \textbf{select\_best}($pop$)\;
    \If{\textbf{fitness}($best$) $<$ \textbf{fitness}($best_{gen}$)}{
        $best \gets best_{gen}$\;
    }
    $gen\gets gen + 1$\;
}
\end{algorithm}

We investigate two ensemble strategies to aggregate scores from individual models. The first, equal weighting, naively sums the scores from each model with uniform weights, providing a straightforward approach. The second, DE-optimised weighting, refines the voting weights using Differential Evolution (DE) algorithm~\cite{storn1995differential}. A distinctive feature of DE, as shown in line 14 of Algorithm~\ref{alg:de}, is its use of scaled differences between agents to guide the generation of new candidates, enabling exploration of the search space. This characteristic has been shown to make DE particularly effective for problems in continuous search spaces~\cite{das2010differential, sohn2023arachne}, making it an appropriate candidate for our weight optimisation task. This optimisation aims to maximise $acc@1$ -- prioritising the buggy method in the top rank -- while minimising wasted effort as a secondary objective in case of ties. With these objectives, we aim to rank the buggy method at top while minimising the number of irrelevant method inspections.

Considering the high inference cost of language models, we further hypothesise that the ensemble approach can better balance the cost-performance trade-off when utilizing multiple models. To enable further analysis, we integrate EnergyMeter~\cite{argerichMeasuringImprovingEnergy2024} to track GPU energy consumption throughout experiments. In addition, we also report the number of tokens and the time taken for the inference as cost of \cosmosfl.

%!TEX root=../paper.tex
\section{Experimental Setup}
\label{sec:experimentalsetup}

\subsection{Research Questions}
Our primary objectives are to evaluate whether the ensemble of small language models enhances fault localisation effectiveness and to investigate the cost-benefit trade-off of ensembles.

\begin{itemize}
    \item \textbf{RQ1. Effectiveness:} To what extent does our ensemble technique improve fault localisation performance? To address this, we conducted initial runs with seven open-source SLMs to assess model orthogonality, running each model five times. Based on these preliminary results, we selected four models that demonstrated the most complementary fault localisation performance in combination. For the final evaluation, we ran AutoFL on each selected model 30 times and sampled a varying number of runs to compare two ensemble weighting strategies: equal and DE-optimised weightings.
    
    \item \textbf{RQ2. Efficiency:} How does the ensemble technique perform in terms of the cost-performance trade-off, and does it contribute to balancing these factors? RQ2 focuses on computational efficiency, considering model inference time, energy consumption, and number of tokens as the measure of cost. We report the sum of input and output tokens as the number of tokens, as this reflects the pricing models of closed-source LLMs more closely.
\end{itemize}

\subsection{Language Models}

To align with the characteristics of smaller language models, we redesign the agent's task from a chat-completion to instruction-following. Each model has been downloaded and served using Ollama~\cite{Ollama2024}, which is chosen for its convenient setup and support for multiple models. We focus on 4-bit quantised models for the sake of memory usage, and choose the following open-source SLMs: \emph{CodeLlama 7B}~\cite{roziereCodeLlamaOpen2024}, \emph{Gemma2 9B}~\cite{teamGemma2Improving2024}, \emph{grantie3 8B}~\cite{Granite30Language2024}, \emph{Llama3 8B}~\cite{dubeyLlama3Herd2024}, \emph{Llama3.1 8B}~\cite{dubeyLlama3Herd2024}, \emph{Mistral NeMo 12B}~\cite{MistralNeMo2024}, and \emph{Qwen2.5-Coder 7B}~\cite{huiQwen25CoderTechnicalReport2024}. These models are all below the size of 8GB (when quantised for 4-bit): we expect them to be compatible with a wider range of machines without GPUs.

Techniques involving inherent randomness require sufficient repetitions to ensure reliable performance measurement~\cite{Arcuri2011ee}. We employ sampling to stabilise our performance metrics and account for stochastic variation of language models. We first run AutoFL with each selected model 30 times. Then, for a given number of runs (R) ranging from 4 to 24, we sample R runs from the 30 runs 20 times for each model. For the case of ensembles, we allocated an equal number of runs to each model, ranging from 1 to 6, resulting in R of multiples of 4.

\subsection{Dataset}

Our evaluation dataset is a subset of Defects4J~\cite{just2014defects4j} used by AutoFL, comprising a total of 353 bugs. We calculate \textit{acc@k} as the number of bugs for which at least one buggy method is correctly ranked within the top \textit{k} places, ensuring consistency with the prior work~\cite{kangQuantitativeQualitativeEvaluation2024a}. Table~\ref{tab:dataset_details} summarises the number of bugs, Lines of Code (LOC) measured using \texttt{cloc}~\cite{adanial_cloc}, and the number of methods and tests for each project.

\begin{table}[htbp]
\caption{Evaluation Dataset Details}
\begin{center}
\resizebox{\columnwidth}{!}{%
\begin{tabular}{lrrrr}
\toprule
Project & \#Bugs & LOC             & \#Methods    & \#Tests    \\ \midrule
Chart   & 26     & 78,564--96,382  & 6,378--8,041 & 1,598--2,201 \\
Closure & 131    & 58,989--104,131 & 4,621--8,700 & 2,692--8,625 \\
Lang    & 64     & 16,593--21,810  & 1,794--2,248 & 1,587--2,265 \\
Math    & 106    & 9,471--84,317   & 1,174--6,015 &   686--3,548  \\
Time    & 26     & 26,589--27,795  & 3,535--3,696 & 3,802--4,054 \\
\bottomrule
\end{tabular}}
\label{tab:dataset_details}
\end{center}
\end{table}

\subsection{Hyperparameters \& Environment}

We utilise DEAP~\cite{DEAP_JMLR2012}, a framework for evolutionary compuation, to implement the differential evolution algorithm. To reduce the over-fitting during the optimisation process, we apply 10-fold cross-validation. We select the DE parameters referring to Storn et al.~\cite{stornUsageDifferentialEvolution1996}, a population size of 40 and 30 generations. To foster exploratory behaviour during the optimisation process, we set high differential weight, 1.5, and also the crossover probability to 0.8.

EnergyMeter~\cite{argerichMeasuringImprovingEnergy2024} is an open-source Python project that measures the energy consumption incurred by the hardware. As we focus on the language model inference cost, we only utilise the GPU energy consumption enabled by \texttt{nvidia-smi}~\cite{developer2021nvidia}.

We conduct all our experiments using the Docker image \texttt{nvidia/cuda:11.3.1-runtime-ubuntu20.04} on a Linux server equipped with 252 GB RAM and 40 Intel Xeon Silver 2.40GHz CPUs. A single NVIDIA GeForce RTX 3090 GPU is utilised to accelerate model inference.

%!TEX root=../paper.tex

\section{Results}
\label{sec:results}

\subsection{RQ1. Effectiveness}

\begin{figure}[htbp]
\centerline{\includegraphics[width=0.75\columnwidth]{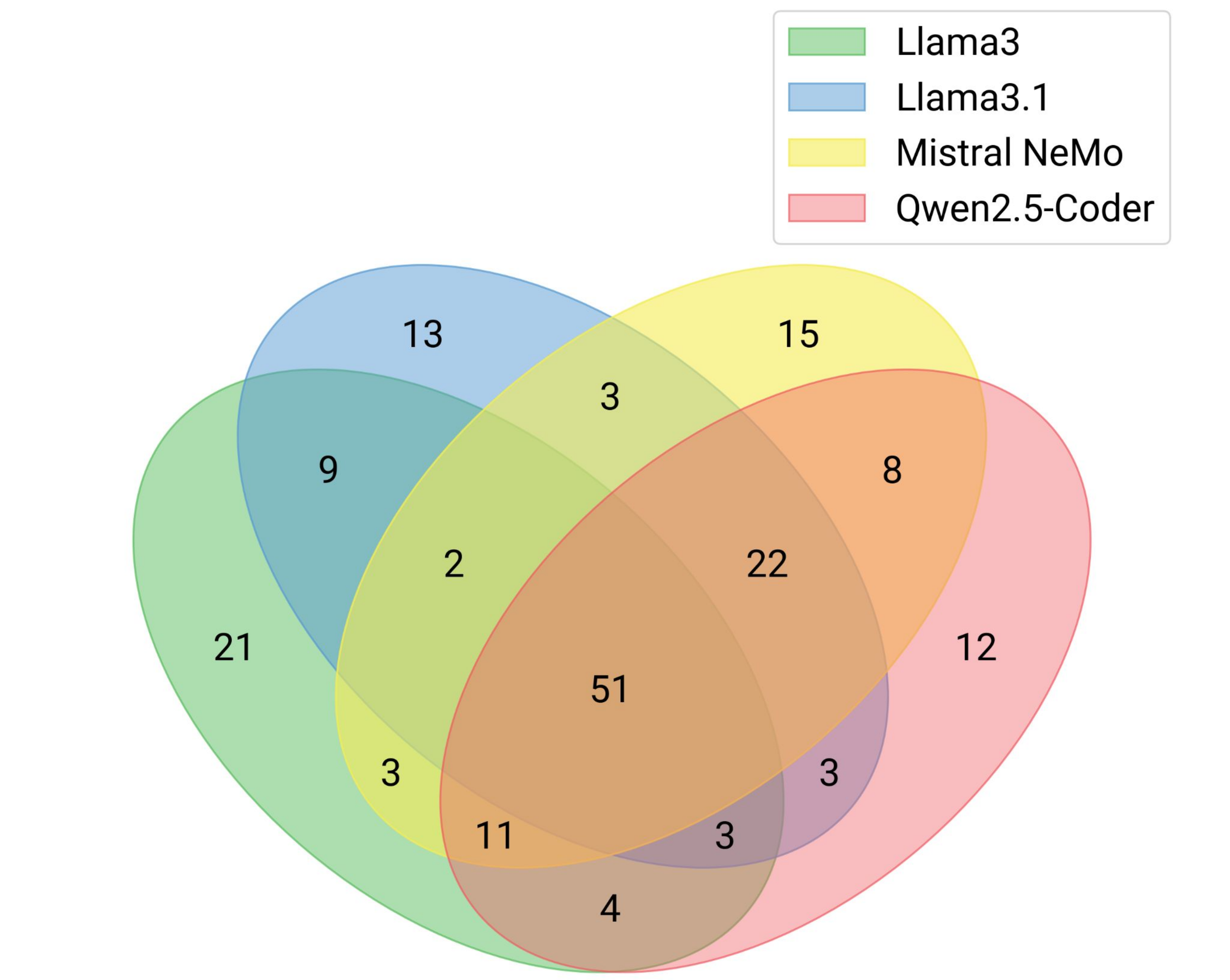}}
\caption{Overlap of bugs ranked at first by Llama3, Llama3.1, Mistral NeMo, and Qwen2.5-Coder. Each model is run 5 times without applying ensemble.}
\label{fig:overlap_of_models}
\end{figure}

The orthogonality among the four selected models—\textit{Llama3 8B}, \textit{Llama3.1 8B}, \textit{Mistral NeMo 12B}, and \textit{Qwen2.5-Coder 7B}—is visualized in Figure~\ref{fig:overlap_of_models}. We identify these models as the most orthogonal combination, collectively ranking 180 bugs first, highlighting their complementary strengths. Building on this, we examine the effectiveness of the ensemble approach for fault localization.

\begin{figure}[htbp]
\centerline{\includegraphics[width=\columnwidth]{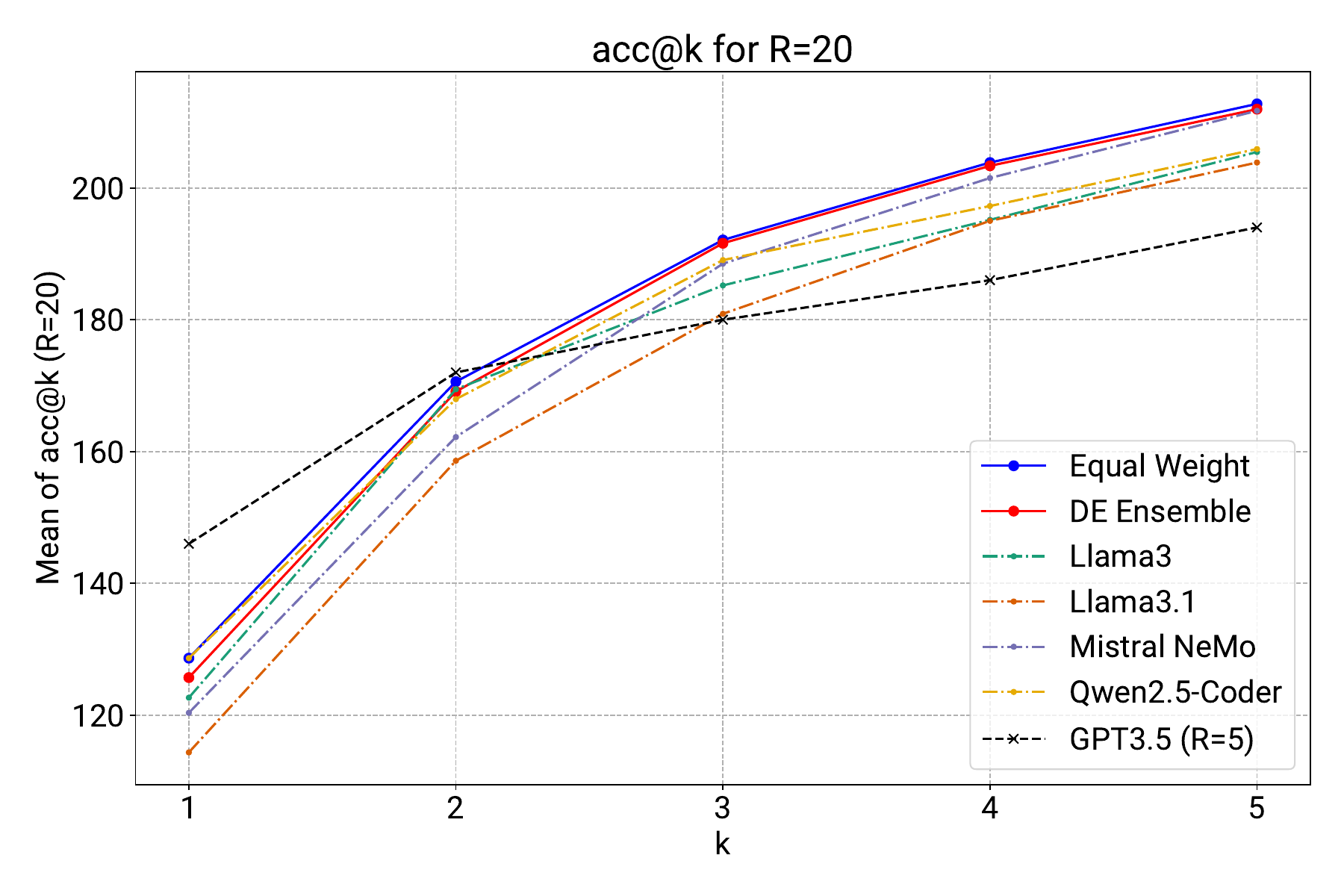}}
\caption{acc@k for R=20 for each model and ensemble approaches, alongside AutoFL’s reported GPT-3.5 performance.}
\label{fig:acc_at_k}
\end{figure}

Figure~\ref{fig:acc_at_k} presents the mean $acc@k$ for $k$ ranging from 1 to 5 across individual models and two ensemble approaches. The results also include the reported performance of GPT-3.5 for $R=5$. Both ensembles tend to outperform individual models as $k$ increases. Since our experiments use more runs ($R=20$) compared to the prior work ($R=5$), a direct comparison is unfair. However, the observation that even the least effective model, Llama3.1, achieves higher accuracies at $k=4$ and 5 suggests that increasing the number of runs could address the underperformance at higher $k$ values relative to SBFL methods, previously attributed to the inability to \textit{dig deep} into a repository~\cite{kangQuantitativeQualitativeEvaluation2024a}.

\begin{figure}[htbp]
\centerline{\includegraphics[width=0.9\columnwidth]{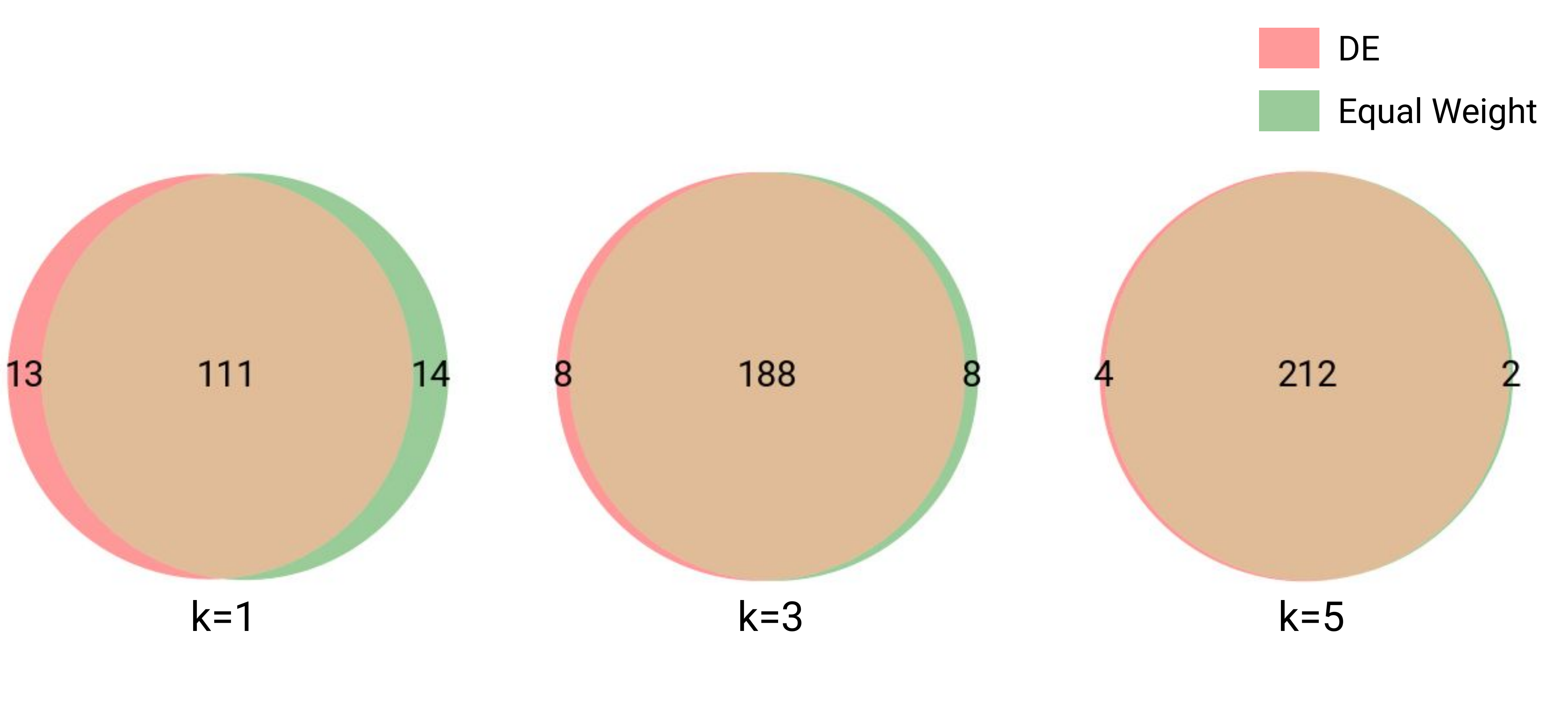}}
\caption{Overlap of top-ranked bugs at each $k$ for a single sample ($R=20$) of DE and Equal Weight Ensembles.}
\label{fig:overlap_at_k}
\end{figure}

Since DE-optimised weights and equal weights show similar performance, we further analyse overlap for a sample of five runs from each model, totalling $R=20$. Figure~\ref{fig:overlap_at_k} indicates that applying different weights for the same runs yields mild variation in bugs ranked at top, suggesting potential for further optimisations to harness model orthogonality. As $k$ increases, however, the top-ranked bug sets become more consistent across weighting methods. SLMs list a limited number of methods as suspicious, allocating confidence score only for those. Thus, when we count the number of methods ranked at top five, they rather show higher consistency; in contrast, for the $acc@1$, they are more sensitive to the weight changes.

\begin{figure}[htbp]
\centerline{\includegraphics[width=\columnwidth]{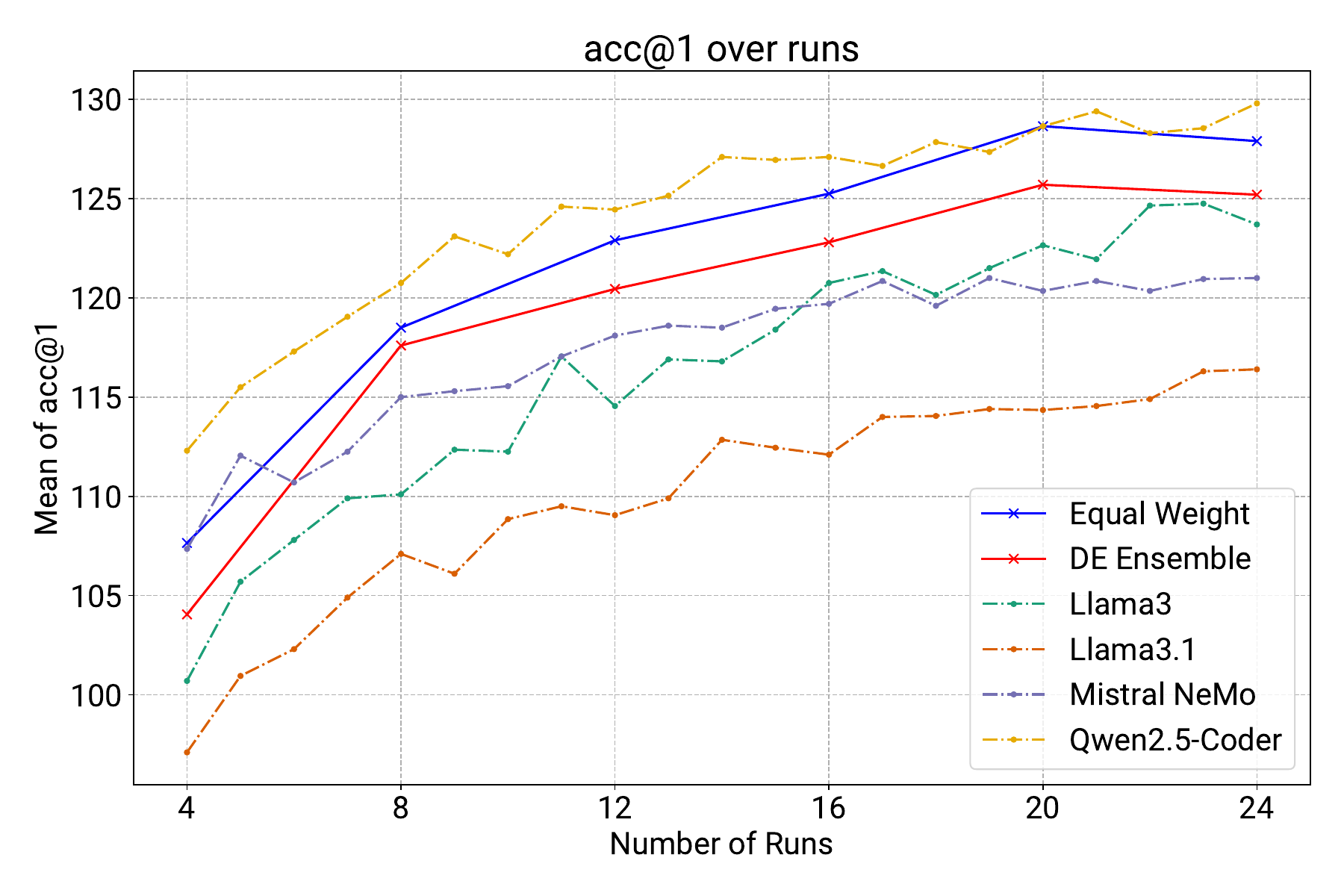}}
\caption{Mean of $acc@1$ across runs for four single models and two ensemble approaches. Note that the ensemble techniques are only available at multiples of four runs.}
\label{fig:acc_at_one_all}
\end{figure}

\begin{figure}[htbp]
\centerline{\includegraphics[width=0.8\columnwidth]{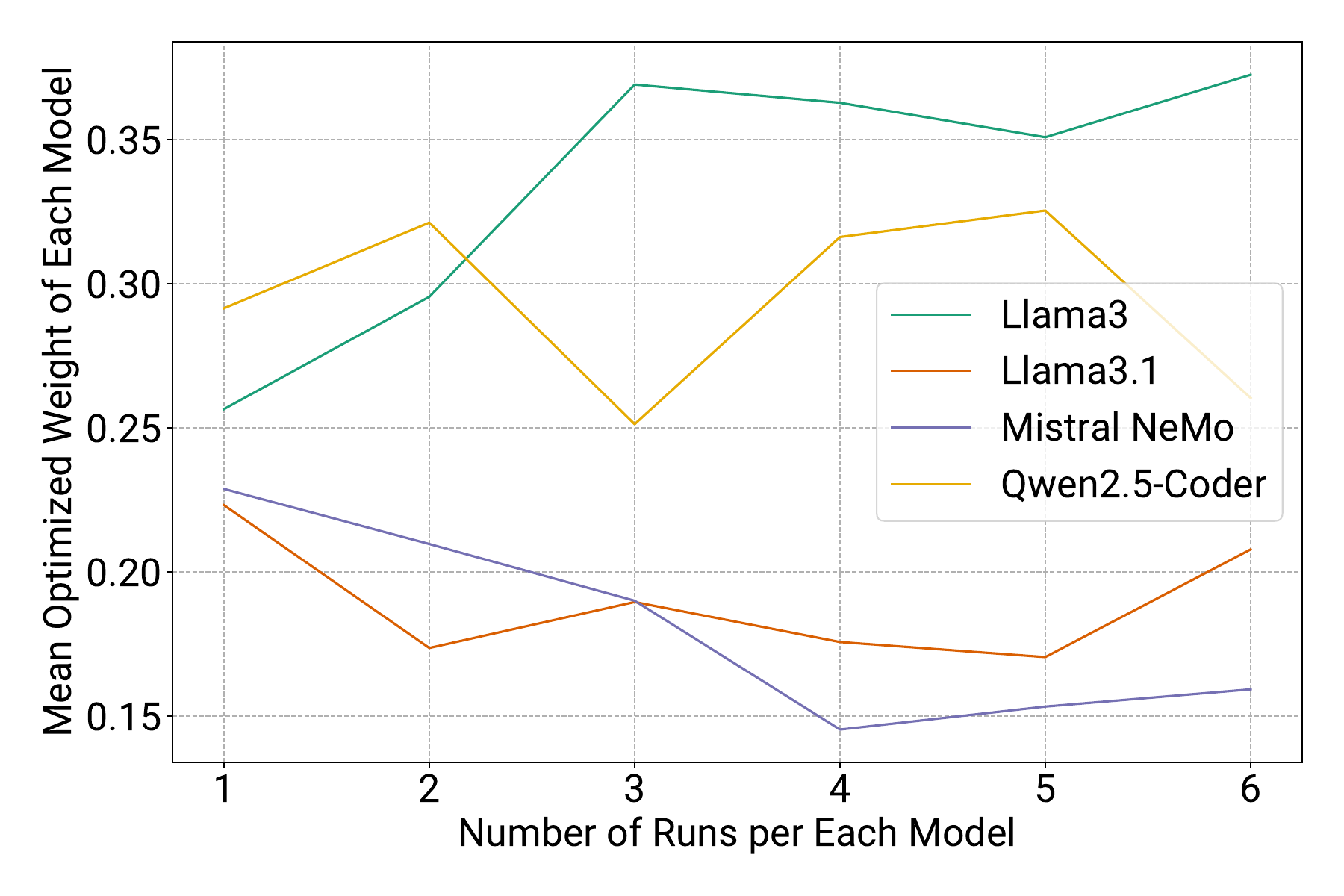}}
\caption{Mean of optimised weights for each model over cross-validation folds and samples.}
\label{fig:mean_best_weights}
\end{figure}

High accuracy at the first rank is crucial for FL tasks, so we focus on $acc@1$ in Figure~\ref{fig:acc_at_one_all}. The ensemble does not surpass the best single model’s performance when given the same number of runs. This outcome can be attributed to how runs are distributed: assigning all weights to a single model effectively limits that model to R/4 runs in the ensemble setup, rather than the full R runs it would receive if evaluated independently. Consequently, mean optimised weights over samples and cross-validation folds, depicted in Figure~\ref{fig:mean_best_weights}, ranged from 0.15 to 0.37 for all four models. Although the DE-based optimisation underperforms the equal weighting, resulting weights still utilize information from all runs.

\begin{figure}[htbp]
\centerline{\includegraphics[width=\columnwidth]{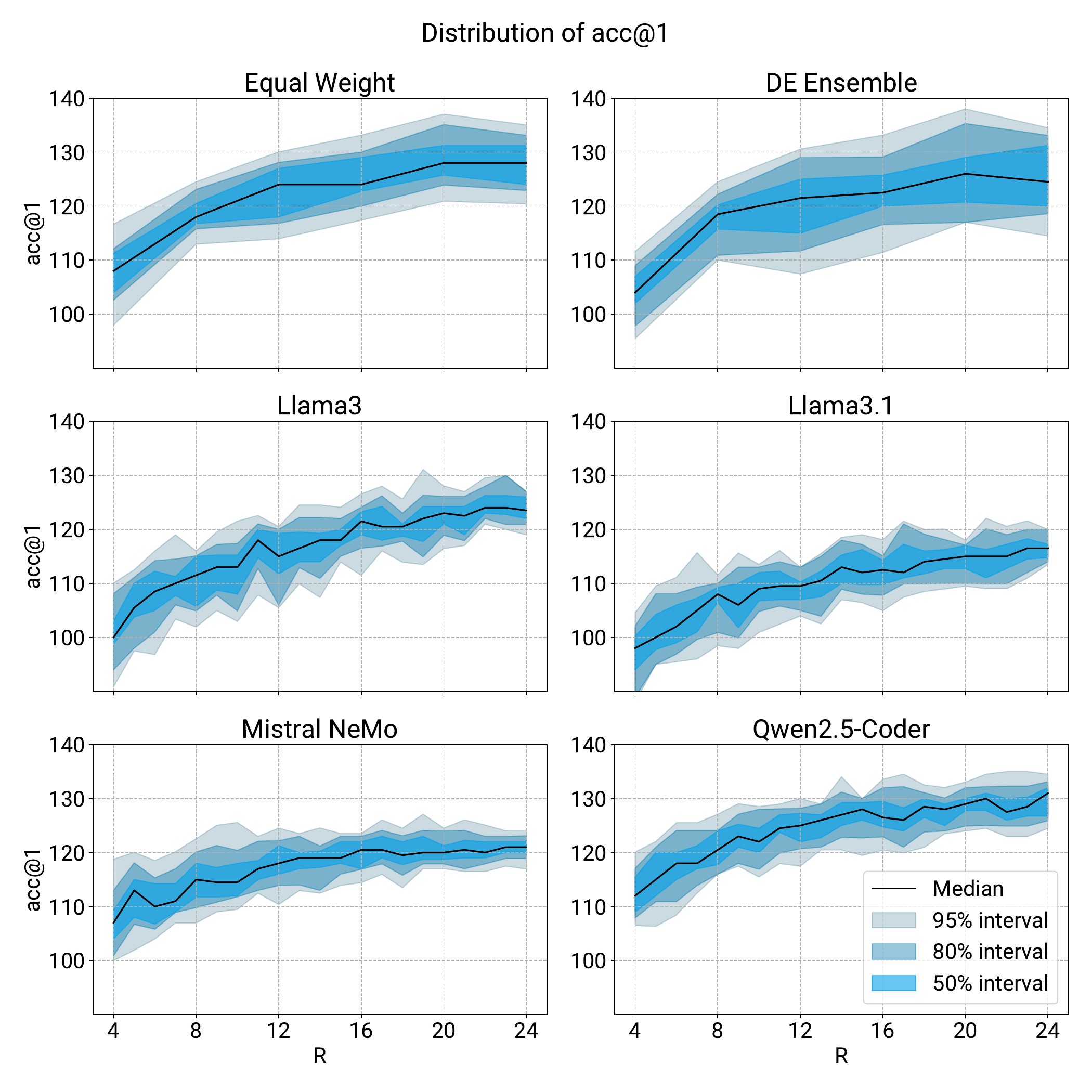}}
\caption{acc@1 across runs for each model and ensemble.}
\label{fig:acc_at_one_each}
\end{figure}

While individual models tend to converge as $R$ increases, ensembles maintain a relatively stable variance level across runs, despite generally outperforming individual models. Figure~\ref{fig:acc_at_one_each} illustrates the distribution of $acc@1$ over samples, capturing this trend. This highlights the need to explore more advanced methods for constructing ensembles.

\begin{figure*}[h]
\centering
\begin{subfigure}[t]{0.32\textwidth}
\includegraphics[width=\textwidth]{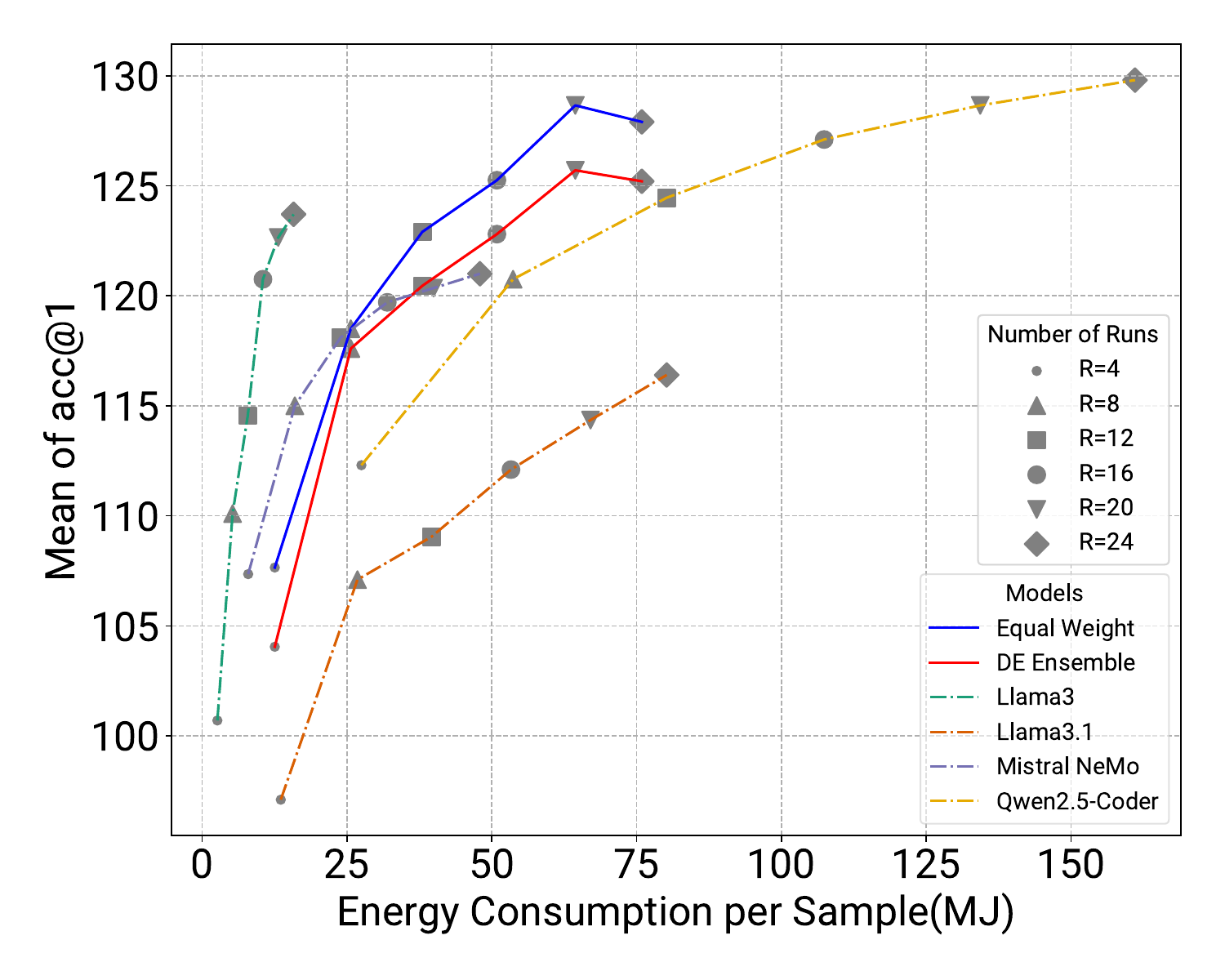}
\caption{Energy Consumption vs. $acc@1$\label{fig:acc_energy}}
\end{subfigure}
\begin{subfigure}[t]{0.32\textwidth}
\includegraphics[width=\textwidth]{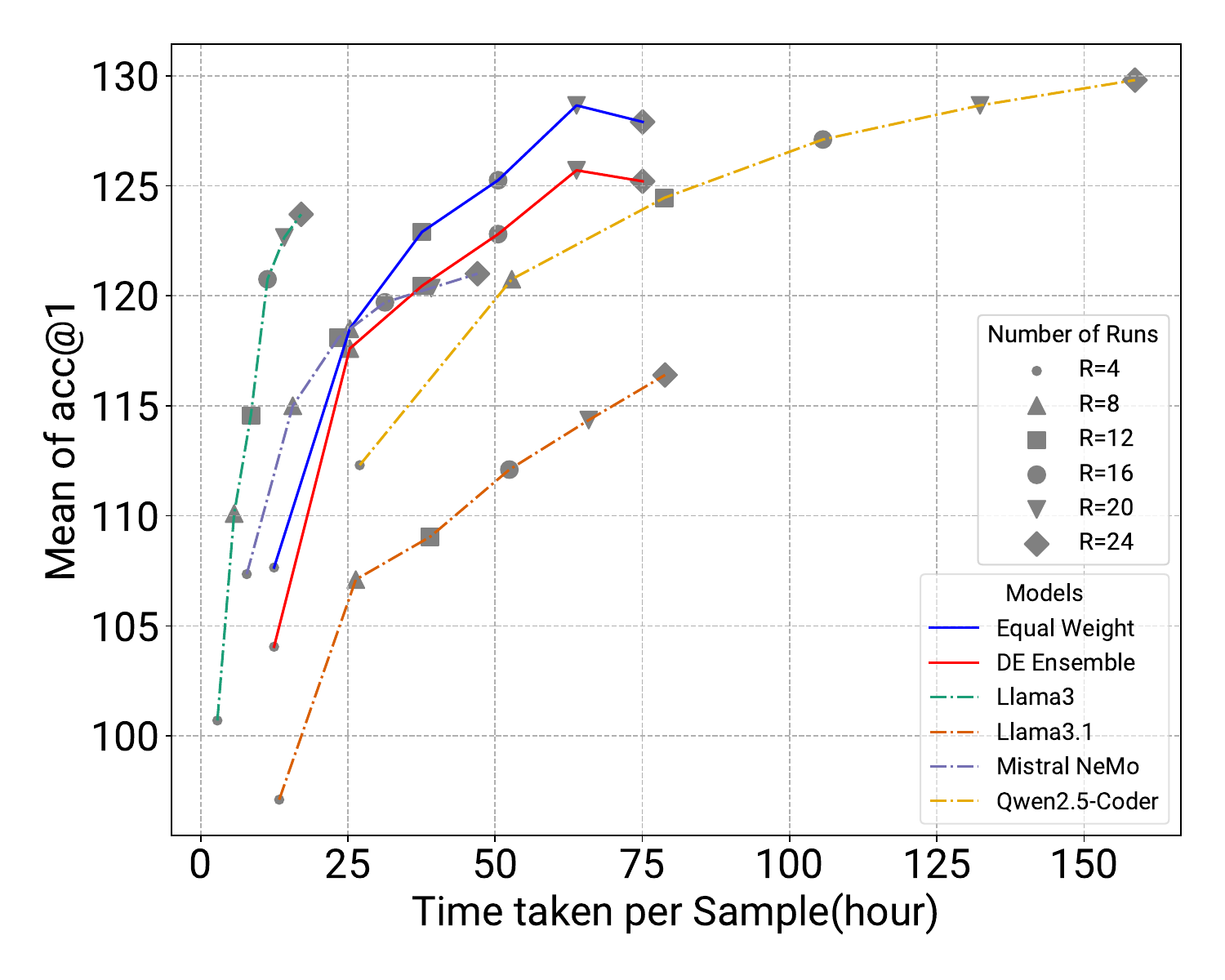}
\caption{Execution Time vs. $acc@1$\label{fig:acc_time}}
\end{subfigure}
\begin{subfigure}[t]{0.32\textwidth}
\includegraphics[width=\textwidth]{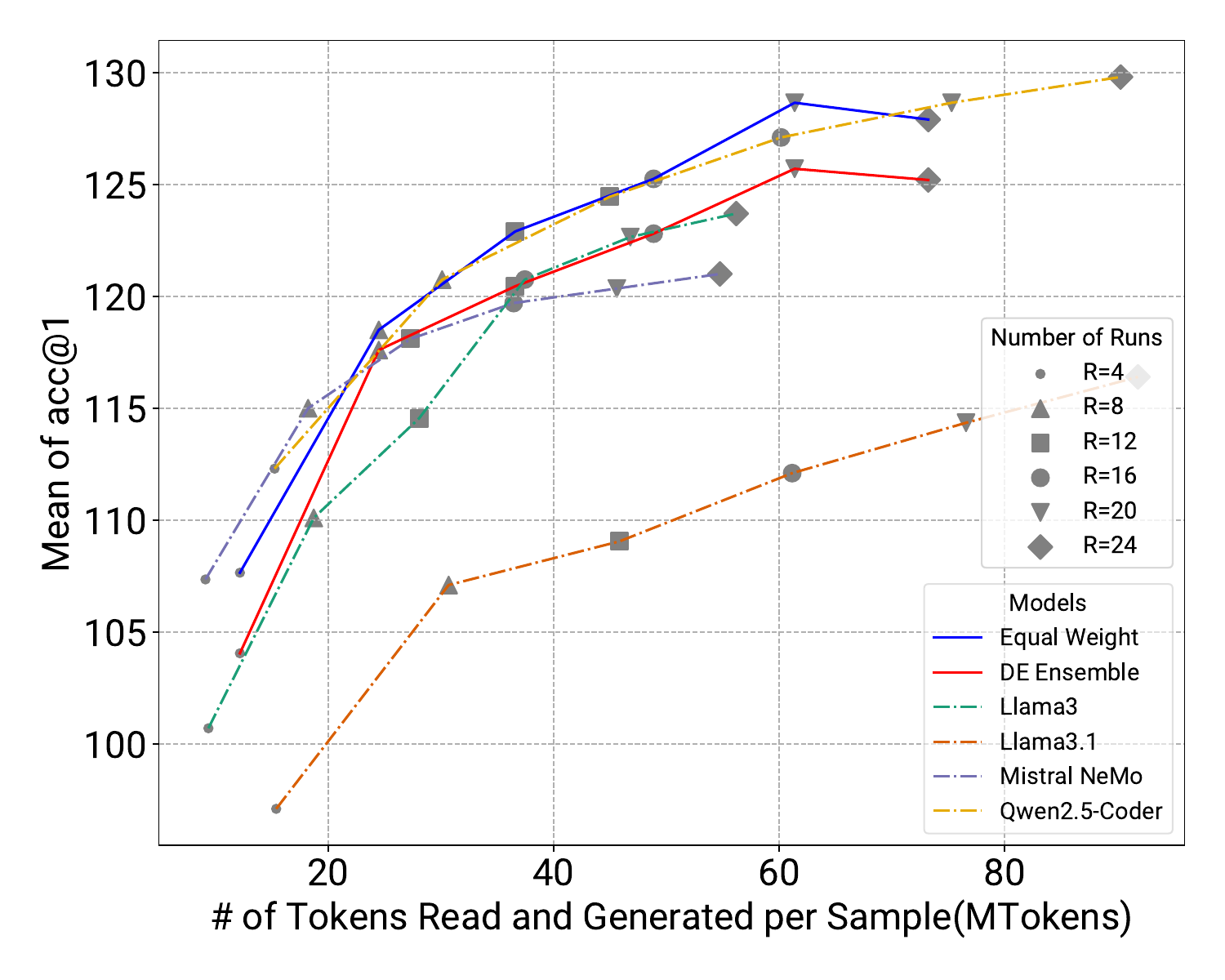}
\caption{\# of Tokens vs. $acc@1$\label{fig:acc_tokens}}
\end{subfigure}
\caption{Cost-Benefit Tradeoff: average energy consumption, inference time, and \# of input and output tokens, across 20 samples}
\end{figure*}

\subsection{RQ2. Efficiency}

% \begin{figure}[htbp]
% \centerline{\includegraphics[width=\columnwidth]{EOL/figs/acc_energy.pdf}}
% \caption{FL Performance over Energy Consumption. We caculated the mean energy consumption over 20 samples, each sample composed of R runs on the 353 bugs.}
% \label{fig:acc_energy}
% \end{figure}

% \begin{figure}[htbp]
% \centerline{\includegraphics[width=\columnwidth]{EOL/figs/acc_time.pdf}}
% \caption{FL Performance over Execution Time. We caculated the mean of execution time per sample.}
% \label{fig:acc_time}
% \end{figure}

% \begin{figure}[htbp]
% \centerline{\includegraphics[width=\columnwidth]{EOL/figs/acc_token.pdf}}
% \caption{FL Performance over Tokens read and generated. We summed up the number of input and output tokens for queries and calculated the mean count per sample.}
% \label{fig:acc_tokens}
% \end{figure}

% \begin{subfigure}[b]{0.15\textwidth}
%         \centering
%         \includegraphics[width=\textwidth]{figures/Figure_lstm_FAA_ROC_AUC.pdf}
%         \caption{LSTM}
%         \label{fig4_lstm_roc}
%     \end{subfigure}

\begin{figure*}[ht]
\centering
\begin{subfigure}[t]{0.32\textwidth}
\includegraphics[width=\textwidth]{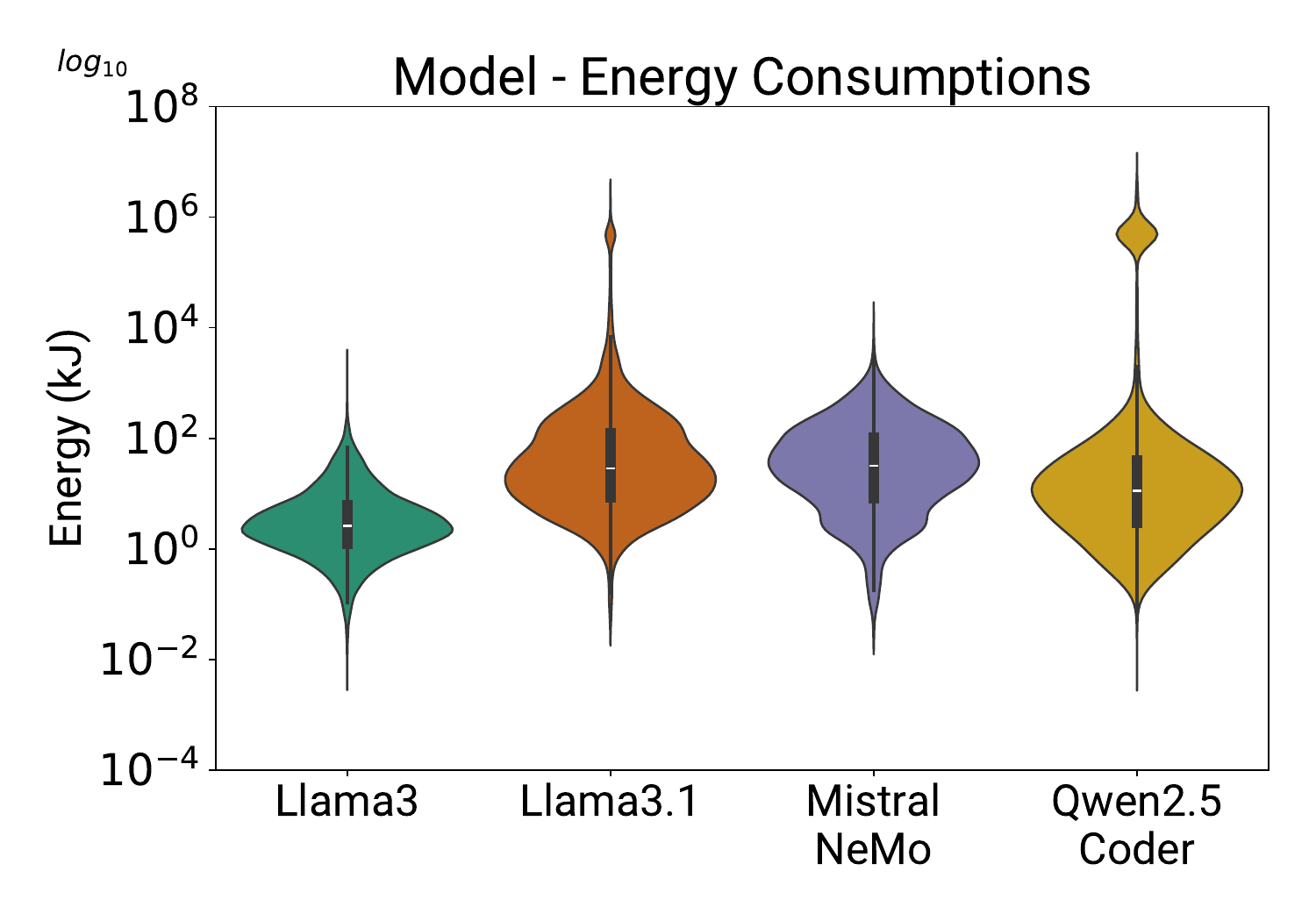}
\caption{Energy consumption per run\label{fig:original_cost}}
\end{subfigure}
\begin{subfigure}[t]{0.32\textwidth}
\includegraphics[width=\textwidth]{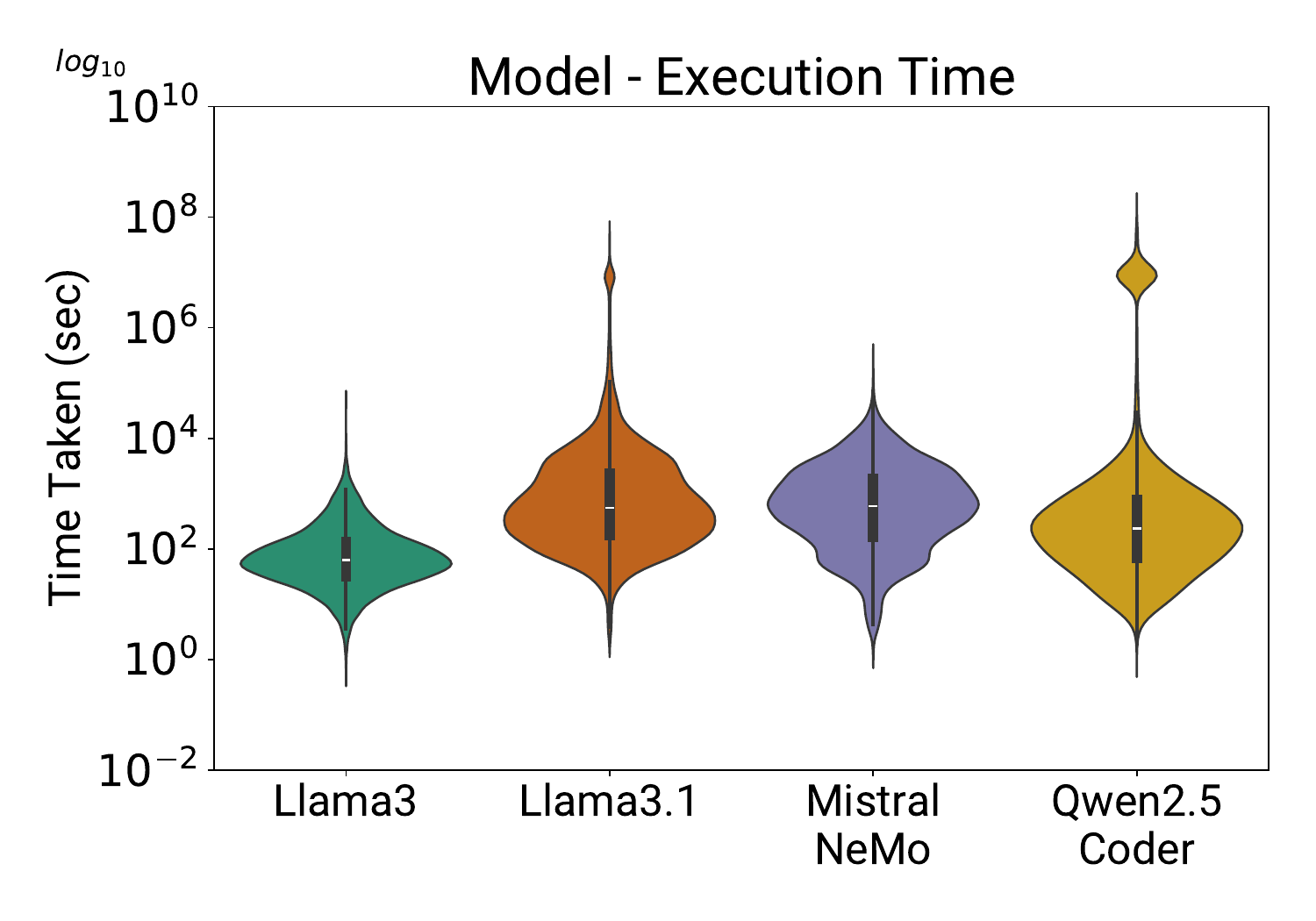}
\caption{Execution time (sec) per run\label{fig:time_cost}}
\end{subfigure}
\begin{subfigure}[t]{0.32\textwidth}
\includegraphics[width=\textwidth]{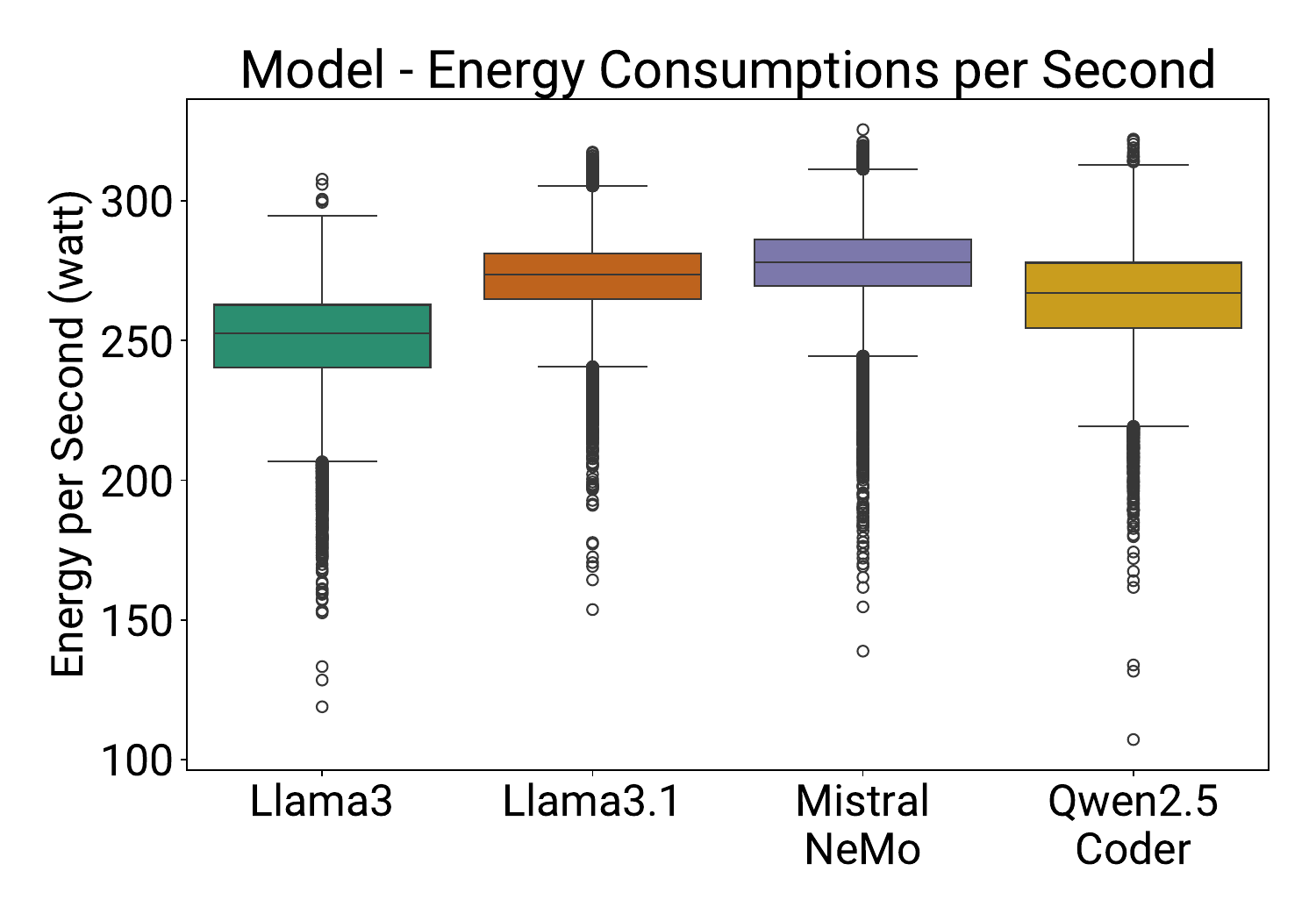}
\caption{Energy consumption per second\label{fig:watt_cost}}
\end{subfigure}
\caption{Boxplots of cost measures per model. Note that $y$-axis of (a) and (b) is log of the measure.}
\end{figure*}

Figures~\ref{fig:acc_energy} to \ref{fig:acc_tokens} show the average cost of each sample, composed of R runs, and its mean $acc@1$. The best performing model, Qwen2.5-Coder consumes the most amount of energy and time, while Llama3 requires the least. Ensemble cost and performance tend to lie between these two extremes. We note that Qwen2.5-Coder is less of an outlier for the number of tokens in Figure~\ref{fig:acc_tokens}, due to the imbalance in its performance. In our evaluation, Qwen2.5-Coder spends significant amounts of time and energy when generating tokens, but input tokens take up the majority of the number of tokens used by an inference run, balancing out the energy and time cost.

We visualize the distribution of energy consumptions and execution time in Figure~\ref{fig:original_cost} and \ref{fig:time_cost} to further investigate the consumption pattern. Unlike the per sample consumptions, all four models exhibit similar median value. The median consumption of Qwen2.5-Coder is actually the second smallest among the four models. However, Llama3.1 and Qwen2.5-Coder show a cluster of outliers at around 10,000 times the median value, which we suspect is linked to an issue in Ollama, given that there is an issue report about some models generating tokens endlessly~\cite{OllamaGetsStuck}. This erroneous behaviour may have inflated Qwen2.5-Coder's overall energy consumption. Since the total energy consumption and execution time follow similar trends, the energy consumption per second remains relatively constant, as shown in Figure~\ref{fig:watt_cost}.

%!TEX root=../paper.tex

\section{Discussion}
\label{sec:discussion}

\begin{figure*}[t]
\centerline{\includegraphics[width=0.95\textwidth]{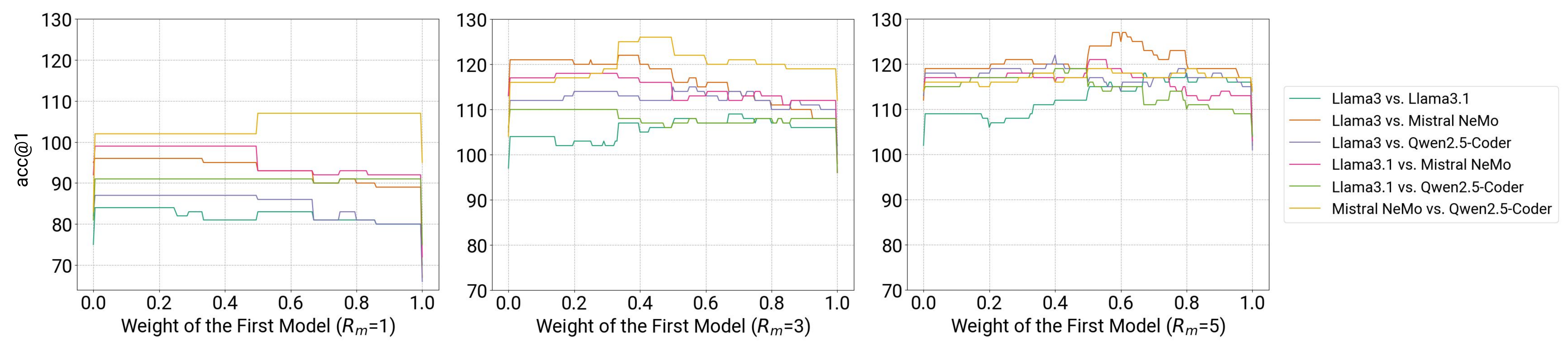}}
\caption{Landscapes of Pairwise $acc@1$ explored by grid search}
\label{fig:grid_landscape}
\end{figure*}

% \fixme{Supplementary for grid search results?}

To better understand the performance characteristics of the two ensemble schemes, we conduct a grid search over weights for pairwise ensembles of models, increasing the number of runs ($R_{m}$) from one to five, as illustrated in Figure~\ref{fig:grid_landscape}. Interestingly, the optimal weights in most scenarios are close to equal, supporting the efficacy of equal weight ensembles. As $R_{m}$ increases, the $acc@1$ landscape becomes more rugged, implying a heightened risk of DE converging to local minima that do not generalise well to validation sets.

We conjecture that introducing a more robust optimisation scheme could further harness the potential of the models. In this context, emerging techniques like expert choice routing~\cite{zhou2022mixture}, which originated from Mixture-of-Experts architectures, present promising avenues. Such approaches could potentially direct computational resources to the most suitable model dynamically, thereby enhancing performance and efficiency. Recently, RouteLLM~\cite{ong2024routellm} exploring router designs for compound language model systems provides early evidence of cutting costs while maintaining capabilities.

In contrast to existing works leveraging multiple language models~\cite{ong2024routellm, shen2024decodecollaboratively}, which focus on model specialization or task delegation, our approach emphasizes orthogonality. Specifically, we aim to construct ensembles where each component possesses comparable yet complementary capabilities, rather than relying on stark performance differences. Exploring the combination of models with substantial capability disparities within \name remains an avenue for future work.

% In addition, we are working to generalize \name to other software engineering tasks. Our ongoing efforts include aggregating results from LLM-based automated program repair techniques, with plans to extend this approach to other language model-driven methodologies.

This study focuses on the quantitative aspect of the FL performance. However, the original work~\cite{kangQuantitativeQualitativeEvaluation2024a} also points that LLMs have the potential to provide rationales for their decisions. Since we now have a diverse set of runs, it would be interesting to see if we can rank generated explanations better based on the larger number of runs sampled from different SLMs, with more practical impact for the developers.

%!TEX root=../paper.tex

\section{Related Work}
\label{sec:relatedwork}

\subsection{LLM-based Fault Localisation}

Fault localisation aims to automatically pinpoint software bugs, often by analysing program behaviour at different granularity levels. With large language models (LLMs) now available, researchers have explored using these models for fault localisation tasks. For instance, Wu et al.~\cite{wuLargeLanguageModels2023} achieved notable results in statement-level localisation given the buggy method using ChatGPT-4, though they identified challenges in extending the model’s context handling to the class level. Addressing this issue, AutoFL~\cite{kangQuantitativeQualitativeEvaluation2024a} introduced an agent architecture that leverages the function-calling capabilities of OpenAI models, allowing LLMs to autonomously explore repositories. This approach also posits that LLMs can potentially offer explanations for root causes of bugs. Similarly, FuseFL~\cite{widyasariDemystifyingFaultyCode2024a} aimed to enhance explainability by integrating spectrum-based fault localisation results into prompts, though its scope was limited to student programming assignments.

Alongside these advancements, efforts to utilize open-source language models in fault localisation have gained traction. Yang et al.~\cite{yangLargeLanguageModels2024a} proposed a test-free FL approach by fine-tuning large, open-source language models with datasets of buggy programs, with a focus on reducing programmer input. More recently, Liu et al.~\cite{liuEmpiricalEvaluationLarge2024} conducted an empirical evaluation comparing FL performance between proprietary and open-source models on novice programming tasks. They found that while ChatGPT-3.5 and 4 outperformed other models, open-source models demonstrated complementary behaviors, successfully localizing bugs that proprietary models missed.

\subsection{Ensemble Methods}

Ensemble learning combines predictions from multiple models to enhance overall performance by leveraging the diverse strengths of individual learners. Ensembles have been applied to fault localisation, based on the observation that no single FL technique is effective across all faults. Wang et al.~\cite{wangSearchbasedFaultLocalization2011} and Xuan et al.~\cite{xuanLearningCombineMultiple2014}, combined FL outputs from multiple models to improve localisation accuracy. Sohn et al.~\cite{sohnWhyTrainselectWhen2019b} further explored ensemble learning across FL methods, demonstrating that combining diverse techniques could achieve better performance. \cosmosfl is the first ensemble technique for LLM-based FL to the best of our knowledge.

Recent works have explored the use of multiple language models in ensemble methods. Jiang et al.~\cite{jiang2023llmblender} proposed an ensemble approach that leverages the complementary strengths of various language models by ranking and fusing their outputs, a method distinct from our approach. More closely, Zhang et al.~\cite{zhangInfeREStepbyStepRegex2023} demonstrated that combining outputs from different regex generators through self-consistency decoding resulted in improved performance. Kumar Dipongkor~\cite{kumardipongkorEnsembleMethodBug2024a} applied ensemble methods to bug triaging using BERT variants, showing that a voting-based ensemble consistently outperformed a stacking-based approach (i.e., training an additional layer on top of pre-trained language models' outputs). However, both of these methods aggregate lower-level outputs, while \cosmosfl employs a voting-based ensemble at the task level by aggregating the FL scores.

%!TEX root=../paper.tex

\section{Conclusion}
\label{sec:conclusion}

We present \name, an ensemble of small language models, and evaluate its instantiation on the fault localisation task, \cosmosfl. We first examine the orthogonal behaviour of SLMs by adopting an LLM-based FL technique to evaluate the feasibility of our approach. Then, by implementing two versions of \cosmosfl -- equal weighting and DE-optimised weighting -- we further assess its performance and cost compared to single-model repetitions. Our results show that \cosmosfl has the potential to outperform a single SLM under cost constraints. Finally, we discuss the current limitations of \cosmosfl, including the issues of the SLM serving platform, and outline future directions, such as improving the performance via routing or more robust optimisation strategy and developing a ranking scheme for explanations.

\balance
\bibliographystyle{ieeetr}
\bibliography{ensemble}

\begin{thebibliography}{10}

\bibitem{Fan2023yu}
A.~Fan, B.~Gokkaya, M.~Harman, M.~Lyubarskiy, S.~Sengupta, S.~Yoo, and J.~M.
  Zhang, ``Large language models for software engineering: Survey and open
  problems,'' in {\em Proceedings of the 45th IEEE/ACM International Conference
  on Software Engineering: Future of Software Engineering}, ICSE-FoSE,
  pp.~31--53, May 2023.

\bibitem{Vaswani2017aa}
A.~Vaswani, N.~Shazeer, N.~Parmar, J.~Uszkoreit, L.~Jones, A.~N. Gomez, L.~u.
  Kaiser, and I.~Polosukhin, ``Attention is all you need,'' in {\em Advances in
  Neural Information Processing Systems} (I.~Guyon, U.~V. Luxburg, S.~Bengio,
  H.~Wallach, R.~Fergus, S.~Vishwanathan, and R.~Garnett, eds.), vol.~30,
  Curran Associates, Inc., 2017.

\bibitem{Wei2024aa}
J.~Wei, X.~Wang, D.~Schuurmans, M.~Bosma, B.~Ichter, F.~Xia, E.~H. Chi, Q.~V.
  Le, and D.~Zhou, ``Chain-of-thought prompting elicits reasoning in large
  language models,'' in {\em Proceedings of the 36th International Conference
  on Neural Information Processing Systems}, NIPS '22, (Red Hook, NY, USA),
  Curran Associates Inc., 2024.

\bibitem{wang2022self}
X.~Wang, J.~Wei, D.~Schuurmans, Q.~Le, E.~Chi, S.~Narang, A.~Chowdhery, and
  D.~Zhou, ``Self-consistency improves chain of thought reasoning in language
  models,'' {\em arXiv preprint arXiv:2203.11171}, 2022.

\bibitem{Yao2022qf}
S.~Yao, J.~Zhao, D.~Yu, N.~Du, I.~Shafran, K.~Narasimhan, and Y.~Cao, ``React:
  Synergizing reasoning and acting in language models,'' in {\em Proceedings of
  the International Conference on Learning Representation}, ICLR 2023, 2023.

\bibitem{Feldt2023ax}
R.~Feldt, S.~Kang, J.~Yoon, and S.~Yoo, ``Towards autonomous testing agents via
  conversational large language models,'' in {\em Proceedings of the 38th
  IEEE/ACM International Conference on Automated Software Engineering (ASE)},
  ASE 2023, pp.~1688--1693, 2023.

\bibitem{Bouzenia2024aa}
I.~Bouzenia, P.~Devanbu, and M.~Pradel, ``Repairagent: An autonomous, llm-based
  agent for program repair,'' 2024.

\bibitem{Yoon2024aa}
J.~Yoon, R.~Feldt, and S.~Yoo, ``Intent-driven mobile gui testing with
  autonomous large language model agents,'' in {\em Proceedings of the 16th
  IEEE International Conference on Software Testing, Verification and
  Validation}, ICST 2024, pp.~129--139, 2024.

\bibitem{Strubell2019aa}
E.~Strubell, A.~Ganesh, and A.~McCallum, ``Energy and policy considerations for
  deep learning in {NLP},'' in {\em Proceedings of the 57th Annual Meeting of
  the Association for Computational Linguistics} (A.~Korhonen, D.~Traum, and
  L.~M{\`a}rquez, eds.), (Florence, Italy), pp.~3645--3650, Association for
  Computational Linguistics, July 2019.

\bibitem{Rillig2023aa}
M.~C. Rillig, M.~{\AA}gerstrand, M.~Bi, K.~A. Gould, and U.~Sauerland, ``Risks
  and benefits of large language models for the environment,'' {\em
  Environmental Science \& Technology}, vol.~57, no.~9, pp.~3464--3466, 2023.

\bibitem{dubeyLlama3Herd2024}
A.~Dubey, A.~Jauhri, A.~Pandey, A.~Kadian, A.~{Al-Dahle}, A.~Letman, and
  et~al., ``The {{Llama}} 3 {{Herd}} of {{Models}},'' Aug. 2024.

\bibitem{teamGemma2Improving2024}
G.~Team, M.~Riviere, S.~Pathak, P.~G. Sessa, C.~Hardin, S.~Bhupatiraju, and
  et~al., ``Gemma 2: {{Improving Open Language Models}} at a {{Practical
  Size}},'' Oct. 2024.

\bibitem{huiQwen25CoderTechnicalReport2024}
B.~Hui, J.~Yang, Z.~Cui, J.~Yang, D.~Liu, L.~Zhang, and et~al.,
  ``Qwen2.5-{{Coder Technical Report}},'' Nov. 2024.

\bibitem{achiam2023gpt}
J.~Achiam, S.~Adler, S.~Agarwal, L.~Ahmad, I.~Akkaya, F.~L. Aleman, D.~Almeida,
  J.~Altenschmidt, S.~Altman, S.~Anadkat, {\em et~al.}, ``{GPT-4 Technical
  Report},'' {\em arXiv preprint arXiv:2303.08774}, 2023.

\bibitem{Kang2024aa}
S.~Kang, J.~Yoon, N.~Askarbekkyzy, and S.~Yoo, ``Evaluating diverse large
  language models for automatic and general bug reproduction,'' {\em IEEE
  Transactions on Software Engineering}, vol.~50, no.~10, pp.~2677--2694, 2024.

\bibitem{Ahmed2023aa}
T.~Ahmed and P.~Devanbu, ``Better patching using llm prompting, via
  self-consistency,'' in {\em 2023 38th IEEE/ACM International Conference on
  Automated Software Engineering (ASE)}, pp.~1742--1746, 2023.

\bibitem{kangQuantitativeQualitativeEvaluation2024a}
S.~Kang, G.~An, and S.~Yoo, ``A {{Quantitative}} and {{Qualitative Evaluation}}
  of {{LLM-Based Explainable Fault Localization}},'' {\em Proc. ACM Softw.
  Eng.}, vol.~1, pp.~64:1424--64:1446, July 2024.

\bibitem{just2014defects4j}
R.~Just, D.~Jalali, and M.~D. Ernst, ``Defects4j: A database of existing faults
  to enable controlled testing studies for java programs,'' in {\em Proceedings
  of the 2014 international symposium on software testing and analysis},
  pp.~437--440, 2014.

\bibitem{storn1995differential}
R.~Storn and K.~Price, ``Differential evolution-a simple and efficient adaptive
  scheme for global optimization over continuous spaces,'' {\em International
  computer science institute}, 1995.

\bibitem{das2010differential}
S.~Das and P.~N. Suganthan, ``Differential evolution: A survey of the
  state-of-the-art,'' {\em IEEE transactions on evolutionary computation},
  vol.~15, no.~1, pp.~4--31, 2010.

\bibitem{sohn2023arachne}
J.~Sohn, S.~Kang, and S.~Yoo, ``Arachne: Search-based repair of deep neural
  networks,'' {\em ACM Transactions on Software Engineering and Methodology},
  vol.~32, no.~4, pp.~1--26, 2023.

\bibitem{argerichMeasuringImprovingEnergy2024}
M.~F. Argerich and M.~{Pati{\~n}o-Mart{\'i}nez}, ``Measuring and {{Improving}}
  the {{Energy Efficiency}} of {{Large Language Models Inference}},'' {\em IEEE
  Access}, vol.~12, pp.~80194--80207, 2024.

\bibitem{Ollama2024}
``Ollama.'' https://github.com/ollama/ollama, Nov. 2024.

\bibitem{roziereCodeLlamaOpen2024}
B.~Rozi{\`e}re, J.~Gehring, F.~Gloeckle, S.~Sootla, I.~Gat, X.~E. Tan, and
  et~al., ``Code {{Llama}}: {{Open Foundation Models}} for {{Code}},'' Jan.
  2024.

\bibitem{Granite30Language2024}
{Granite Team, IBM}, ``Granite 3.0 {{Language Models}}.''
  https://github.com/ibm-granite/granite-3.0-language-models/blob/main/paper.pdf,
  2024.

\bibitem{MistralNeMo2024}
{Mistral AI}, ``Mistral {{NeMo}}.'' https://mistral.ai/news/mistral-nemo/, July
  2024.

\bibitem{Arcuri2011ee}
A.~Arcuri and L.~Briand, ``A practical guide for using statistical tests to
  assess randomized algorithms in software engineering,'' in {\em Proceedings
  of the 33rd International Conference on Software Engineering}, ICSE '11,
  pp.~1--10, ACM, 2011.

\bibitem{adanial_cloc}
A.~Danial, ``cloc: v1.92,'' Dec. 2021.

\bibitem{DEAP_JMLR2012}
F.-A. Fortin, F.-M. {De Rainville}, M.-A. Gardner, M.~Parizeau, and C.~Gagn\'e,
  ``{DEAP}: Evolutionary algorithms made easy,'' {\em Journal of Machine
  Learning Research}, vol.~13, pp.~2171--2175, jul 2012.

\bibitem{stornUsageDifferentialEvolution1996}
R.~Storn, ``On the usage of differential evolution for function optimization,''
  in {\em Proceedings of {{North American Fuzzy Information Processing}}},
  pp.~519--523, June 1996.

\bibitem{developer2021nvidia}
N.~Developer, ``Nvidia system management interface,'' {\em NVIDIA System
  Management Interface}, 2021.

\bibitem{OllamaGetsStuck}
``Ollama gets stuck in an infinite loop sometimes and has to be restarted
  {$\cdot$} {{Issue}} \#2805 {$\cdot$} ollama/ollama.''
  https://github.com/ollama/ollama/issues/2805.

\bibitem{zhou2022mixture}
Y.~Zhou, T.~Lei, H.~Liu, N.~Du, Y.~Huang, V.~Zhao, A.~M. Dai, Q.~V. Le,
  J.~Laudon, {\em et~al.}, ``Mixture-of-experts with expert choice routing,''
  {\em Advances in Neural Information Processing Systems}, vol.~35,
  pp.~7103--7114, 2022.

\bibitem{ong2024routellm}
I.~Ong, A.~Almahairi, V.~Wu, W.-L. Chiang, T.~Wu, J.~E. Gonzalez, M.~W. Kadous,
  and I.~Stoica, ``Routellm: Learning to route llms with preference data,''
  {\em arXiv preprint arXiv:2406.18665}, 2024.

\bibitem{shen2024decodecollaboratively}
S.~Z. Shen, H.~Lang, B.~Wang, Y.~Kim, and D.~Sontag, ``Learning to decode
  collaboratively with multiple language models,'' {\em arXiv preprint
  arXiv:2403.03870}, 2024.

\bibitem{wuLargeLanguageModels2023}
Y.~Wu, Z.~Li, J.~M. Zhang, M.~Papadakis, M.~Harman, and Y.~Liu, ``Large
  {{Language Models}} in {{Fault Localisation}},'' Oct. 2023.

\bibitem{widyasariDemystifyingFaultyCode2024a}
R.~Widyasari, J.~W. Ang, T.~G. Nguyen, N.~Sharma, and D.~Lo, ``Demystifying
  {{Faulty Code}}: {{Step-by-Step Reasoning}} for {{Explainable Fault
  Localization}},'' in {\em {{IEEE International Conference}} on {{Software
  Analysis}}, {{Evolution}} and {{Reengineering}}}, pp.~568--579, Mar. 2024.

\bibitem{yangLargeLanguageModels2024a}
A.~Z.~H. Yang, C.~Le~Goues, R.~Martins, and V.~Hellendoorn, ``Large {{Language
  Models}} for {{Test-Free Fault Localization}},'' in {\em Proceedings of the
  {{IEEE}}/{{ACM}} 46th {{International Conference}} on {{Software
  Engineering}}}, {{ICSE}} '24, (New York, NY, USA), pp.~1--12, Association for
  Computing Machinery, Feb. 2024.

\bibitem{liuEmpiricalEvaluationLarge2024}
Y.~Liu, H.~Liu, Z.~Yang, Z.~Li, and Y.~Liu, ``Empirical {{Evaluation}} of
  {{Large Language Models}} for {{Novice Program Fault Localization}},'' in
  {\em 2024 {{IEEE}} 24th {{International Conference}} on {{Software Quality}},
  {{Reliability}} and {{Security}} ({{QRS}})}, pp.~180--191, July 2024.

\bibitem{wangSearchbasedFaultLocalization2011}
S.~Wang, D.~Lo, L.~Jiang, {Lucia}, and H.~C. Lau, ``Search-based fault
  localization,'' in {\em 2011 26th {{IEEE}}/{{ACM International Conference}}
  on {{Automated Software Engineering}} ({{ASE}} 2011)}, pp.~556--559, Nov.
  2011.

\bibitem{xuanLearningCombineMultiple2014}
J.~Xuan and M.~Monperrus, ``Learning to {{Combine Multiple Ranking Metrics}}
  for {{Fault Localization}},'' in {\em 2014 {{IEEE International Conference}}
  on {{Software Maintenance}} and {{Evolution}}}, pp.~191--200, Sept. 2014.

\bibitem{sohnWhyTrainselectWhen2019b}
J.~Sohn and S.~Yoo, ``Why train-and-select when you can use them all? ensemble
  model for fault localisation,'' in {\em Proceedings of the {{Genetic}} and
  {{Evolutionary Computation Conference}}}, {{GECCO}} '19, (New York, NY, USA),
  pp.~1408--1416, Association for Computing Machinery, July 2019.

\bibitem{jiang2023llmblender}
D.~Jiang, X.~Ren, and B.~Y. Lin, ``Llm-blender: Ensembling large language
  models with pairwise ranking and generative fusion,'' {\em arXiv preprint
  arXiv:2306.02561}, 2023.

\bibitem{zhangInfeREStepbyStepRegex2023}
S.~Zhang, X.~Gu, Y.~Chen, and B.~Shen, ``{{InfeRE}}: {{Step-by-Step Regex
  Generation}} via {{Chain}} of {{Inference}},'' in {\em 2023 38th
  {{IEEE}}/{{ACM International Conference}} on {{Automated Software
  Engineering}} ({{ASE}})}, pp.~1505--1515, Sept. 2023.

\bibitem{kumardipongkorEnsembleMethodBug2024a}
A.~Kumar~Dipongkor, ``An {{Ensemble Method}} for {{Bug Triaging}} using {{Large
  Language Models}},'' in {\em Proceedings of the 2024 {{IEEE}}/{{ACM}} 46th
  {{International Conference}} on {{Software Engineering}}: {{Companion
  Proceedings}}}, {{ICSE-Companion}} '24, (New York, NY, USA), pp.~438--440,
  Association for Computing Machinery, May 2024.

\end{thebibliography}

\end{document}